\newcounter{myctr}
\def\myitem{\refstepcounter{myctr}\bibfont\noindent\ifnum\themyctr>9\else\phantom{0}\fi\hangindent17pt\themyctr.\enskip}

\documentclass{ws-ijqi}
\usepackage{hyperref}
\usepackage[super,sort,compress]{cite}

\usepackage{float}
\usepackage{multirow}
\usepackage[T1]{fontenc}
\usepackage{amsfonts}
\usepackage{amsmath}    
\usepackage{amssymb}
\usepackage{graphicx}   
\usepackage{dsfont}
\usepackage{subfigure}  
\usepackage{braket}
\usepackage{xcolor}

\begin{document}
\title{Minimal scenario facet Bell inequalities for
multi-qubit states}

\author{Arpan Das \footnote{Present address: Institute of Physics, Faculty of Physics, Astronomy and Informatics, Nicolaus Copernicus University, Grudzi\k{a}dzka 5/7, 87-100
Toru\'{n}, Poland} }

\address {Institute of Physics, Sachivalaya Marg, Sainik School, Bhubaneswar 751005, Odisha, India\\Homi Bhabha National Institute, Training School Complex, Anushakti Nagar, Mumbai 400085, India.\\
arpand@umk.pl}

\author{ Chandan Datta \footnote{Present address: Centre of New Technologies, University of Warsaw, Banacha 2c, 02-097 Warsaw, Poland.}}

\address{Institute of Physics, Sachivalaya Marg, Sainik School, Bhubaneswar 751005, Odisha, India.\\Homi Bhabha National Institute, Training School Complex, Anushakti Nagar, Mumbai 400085, India.\\
chandan@iopb.res.in} 
\author{Pankaj Agrawal} 
\address{ Institute of Physics, Sachivalaya Marg, Sainik School, Bhubaneswar 751005, Odisha, India,\\Homi Bhabha National Institute, Training School Complex, Anushakti Nagar, Mumbai 400085, India.\\
agrawal@iopb.res.in}


\maketitle

\begin{abstract}
Facet inequalities play an important role in detecting the nonlocality of a quantum state. 
The number of such inequalities depends on the Bell test scenario. With the increase in 
the number of parties, measurement outcomes, or/and the number of measurement settings, 
there are more nontrivial facet inequalities. For several Bell scenarios, involving two 
dichotomic measurement settings for two parties and one dichotomic measurement by other parties,  we show that the local polytope has only one non-trivial facet. For three parties, we have three variants of this inequality, depending upon which party is doing one dichotomic measurement. This measurement scenario for a multipartite
state may be considered as the minimal scenario involving multipartite correlations that can
detect nonlocality. We show that this inequality is violated by all 
generalized GHZ states. Being the only facet Bell inequality, this inequality is also violated by any entangled three-qubit pure state. We also show that for noisy W states, our inequality is more effective than the well-known Mermin inequality. 
\end{abstract}

\section{Introduction}
Bell nonlocality \cite{bell1964}, an intriguing feature of quantum mechanics, has been studied extensively since the time of John S. Bell. Since the introduction of the famous EPR paradox \cite{EPR}, entanglement has been known to be a source of many fascinating phenomena, including Bell nonlocality. However, entanglement in a state does not always guarantee Bell nonlocality; a simple example is the Werner state \cite{werner}. The set of quantum correlations is convex, but they do not form a polytope; whereas the set of local correlations is convex and the correlations form a polytope \cite{brunnerRMP}. The nontrivial facets of this local polytope are known as tight Bell inequalities. The well-known Clauser-Horne-Shimony-Holt (CHSH) inequality \cite{chsh} is an example of facet Bell inequality for two parties, two measurement settings, and two outcomes per setting. It is the only nontrivial facet inequality for this scenario giving the maximal quantum violation of $2\sqrt{2}$ which is also Tsirelson's bound \cite{tsirelson}. For three parties and two dichotomic measurement settings per party, {\'S}liwa \cite{sli} constructed the local polytope where Mermin inequality \cite{mermin}  is one of the facets.  Originally, Mermin inequality was proposed for $n$-qubit GHZ state, where all the operators are either $\sigma_x$ or $\sigma_y$. Ardehali inequality \cite{ardehali} was different from Mermin’s in the $n$th operator, where any arbitrary measurement settings $a$ and $b$ are allowed. Belinskii and Klyshko \cite{bk} first introduced the polynomial
versions of inequalities. Nowadays all these inequalities are together called Mermin-Ardehali-Belinskii-Klyshko (MABK) inequalities  \cite{mermin,ardehali,gisinp}. In our previous paper \cite{arpan}, we noted a particular limitation of MABK inequalities.
Particularly, the $n$-qubit state, $\ket{\psi}=\cos\alpha \ket{0...0}+\sin\alpha \ket{1...1}$ (generalized GHZ state) does not violate MABK inequalities \cite{gen} for $\sin 2\alpha \le 1/\sqrt{2^{n-1}}$. Not only that in \cite{all} all correlation Bell inequalities were constructed for $n$ qubits and in \cite{nonvio} it was shown that generalized GHZ states do not violate those inequalities for the whole parameter range provided one is restricted to rank-1 qubit measurements. In fact, in Ref \cite{degenerate}, authors showed that if one allows degenerate qubit measurements, at least for $n=2,3,4$ and $5$, full correlation Bell inequalities get violated for the whole parameter range. 
One can mend this drawback by considering alternative Bell inequalities designed for different settings.

 In \cite{moremeasure} authors have constructed multipartite tight Bell inequalities for more than two measurement settings per party. In this extended scenario, they showed that generalized GHZ states now violate the inequalities for the whole parameter range. But if we restrict ourselves to the scenario where maximum of two measurement settings are allowed per party, then can we construct Bell inequalities, for which this shortcoming can be avoided?
This was the main motivation behind our previous paper \cite{arpan} where we constructed a set of six inequalities each of which is violated by generalized GHZ states for the whole parameter range. Also, with the help of this set of inequalities, we can distinguish between pure biseparable and pure genuinely entangled states. This distinction cannot be done with MABK inequalities, as they give only sufficient criteria \cite{gisinp,uffnik,nsvet,nagata1,yu1} to distinguish them. It is noteworthy that Ref. \cite{nagata1} and Ref. \cite{yu1} give better characterization for this sufficient criteria.
These six inequalities could be obtained from two inequalities after permutations of qubits. One important fact about these inequalities was the scenario we considered, i.e.,  three parties, two dichotomic measurement settings for two parties and one dichotomic measurement for the remaining. For brevity, the scenario can be described by the notation $\{[2 ~2] ~[2~ 2] ~[2]\}$ following the Ref. \cite{lianghidden}, where number of square brackets denotes the number of parties, number of entries in a square bracket stands for the number of inputs for that party and the actual value of the entries stands for the number of outputs for that particular input. But our inequalities were not facet inequalities for this particular scenario. 
 In the multipartite setting, constructing Bell polytope is a non-trivial task. It is not only computationally hard but also there is complicated relationship between multipartite entanglement and multipartite nonlocality, even for qubits. There are variety of local polytopes one can construct for different forms of nonlocality and there are different forms of entanglement as well. Consequently, it is hard to find optimal Bell inequality/inequalities, which will guarantee a given form of entanglement. As facet inequalities are the tightest version of Bell inequalities, if found, it would be the minimal requirement to certify the given form of entanglement.
Question naturally arises what about the facet inequalities for this scenario. Will they also circumvent the obstacle posed by the MABK inequalities regarding the violation in the whole parameter range for generalized GHZ states and order them according to their entanglement? Besides, construction of facet Bell inequalities in this scenario is itself very interesting, as it is the {\it minimal scenario}, where one can generate facet Bell inequalities. 
We need minimum of two parties performing two dichotomic measurements, to have some nontrivial facet inequalities, 
also called facet Bell inequalities.

In this paper, we first explicitly construct the facets of the local polytope for three parties and find only one nontrivial facet inequality up to the relabelling \cite{collins2} of indices. With the permutation of parties, the number is three. Interestingly, this facet inequality is equivalent to the {\it lifted} version \cite{pironio_facet} of CHSH inequality for more than two parties. {
Also in Ref. \cite{sli}, it can be seen that for three parties and two dichotomic measurements per party, there is only one class of facet Bell inequality, where one party makes only one dichotomic measurements.
{\it This shows that to uncover the nonlocality of a three-qubit, or
multiqubit (as discussed below) system, one facet Bell inequality, and its permutations, may be enough.} 
This inequality involves multipartite correlations; so it explores multipartite nonlocality.
We also compare the effectiveness of our inequality with other well-known inequalities. We  also consider a few noisy mixed states and show that for noisy W states, our inequality is more effective than the famous Mermin inequality. 
We also construct the facets of the local polytopes for four and five parties for the same minimal measurement scenario where, only two parties perform two dichotomic measurements and the remaining parties perform one dichotomic measurement each, specifically, for the scenarios $\{[2 ~2] ~[2~ 2] ~[2]~[2]\}$ and $\{[2 ~2] ~[2~ 2] ~[2]~[2]~[2]\}$ respectively. For each of these cases, there is again only one non-trivial facet, up to the relabelling of indices, with a similar structure as the 
three-party scenario. This observation enables us to generalize our facet Bell inequality to $n$-qubit systems. We show that generalized GHZ states of $n$ qubits violate the facet inequality for the whole parameter range. 
Interestingly, the facet Bell inequality we obtain is not maximally violated by a maximally entangled state. The notion of a maximally entangled state for a multipartite state is not straightforward.
However, for a three-qubit system, GHZ-state, for all practical purposes, can be considered to be maximally entangled. We find that the facet Bell inequality of our scenario is not maximally violated by the GHZ-state.

The paper is organized as follows.  In Sec. 2, we obtain facet Bell inequalities in the
case of three qubits for our minimal scenario. Section 3 deals with three-qubit generalized GHZ states. Section 4 deals with mixed state scenario. Section 5 deals with
extension to multipartite solution. Section 6 concludes.

\section{Facet Inequalities}
Before stating our results, we briefly review the polytope formed by local correlations and 
the significance of facet Bell inequalities. Polytope is a generalization of polygons to any dimension.
 Mathematically, there are two equivalent definitions \cite{polytope} of a polytope: $V$ representation and $H$ representation. A $V$-polytope is the convex hull of a finite set of points $\in\mathbb{R}^d$, which are called vertices. A $H$ polytope is an intersection of a finite number of closed half-spaces in some $\mathbb{R}^d$, which is bounded. So, a polytope is a set of points $P\subseteq \mathbb{R}^d$, which can be represented as either a $V$ or a $H$ polytope. The dimension of a polytope is the dimension of its affine hull. 
A Bell experiment can be described as follows. 
A source $S$ distributes two particles (which may be entangled) to two spatially separated parties, Alice (A) and Bob (B). This situation can be easily generalized to multipartite scenarios, but for simplicity, we are discussing the preliminaries for two parties only. Now, Alice and Bob make local measurements labeled by the inputs $x$ and $y$ respectively. The outputs of their measurements are given by $a$ and $b$.
\begin{figure}[h]
\begin{center}
\includegraphics[scale=0.28]{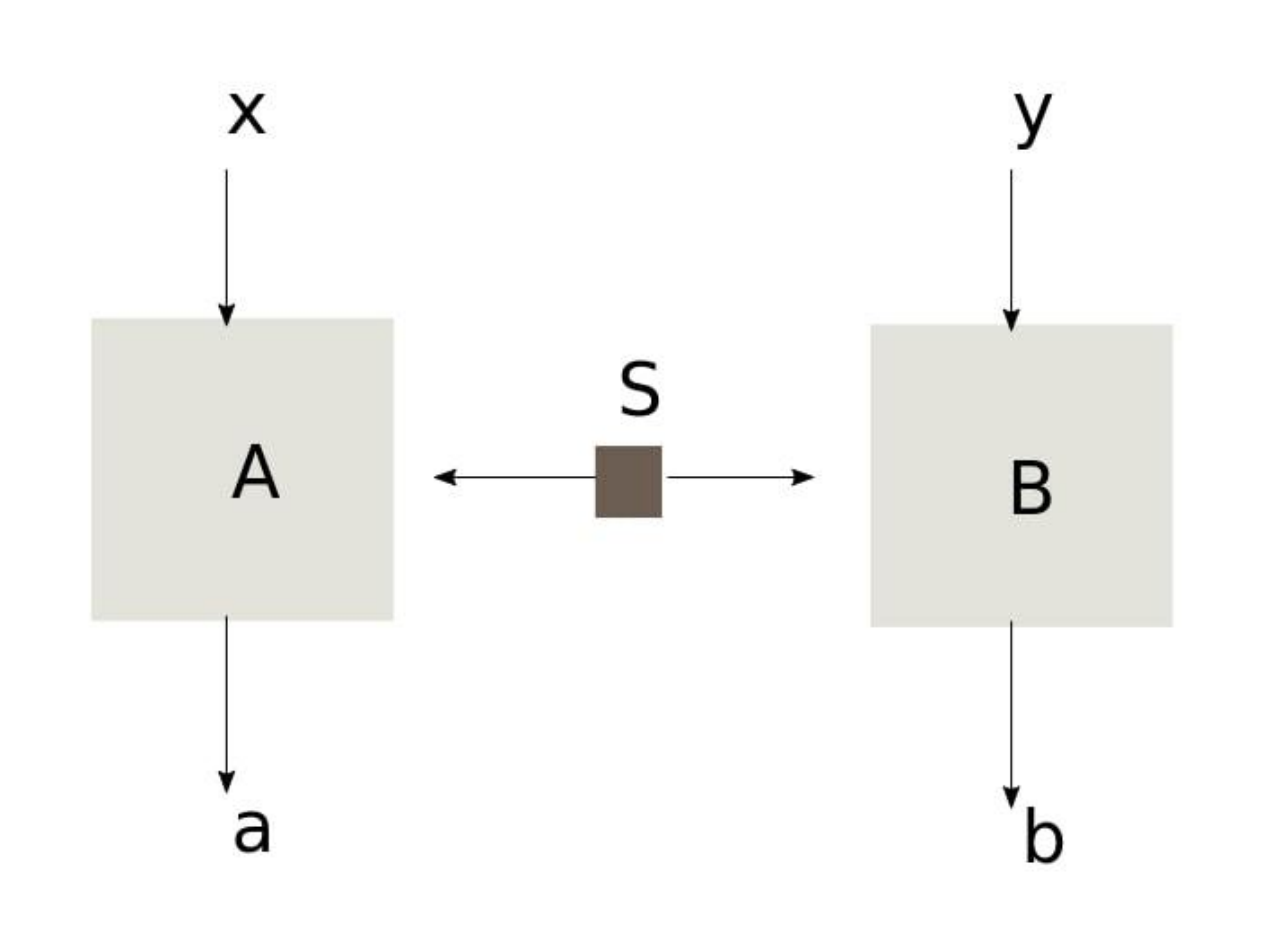}
\caption{ Bell experiment with two parties.}
\end{center}
\end{figure}
The joint probability distribution $\textbf{p}=\{p(ab|xy)\}$ characterizing the Bell experiment is called correlations or behavior. We are interested only in these correlations, anything else is a black-box. Local Causality (LC) or Factorizability or Bell locality is defined as -- 
$p(ab|xy)=\int_\lambda d\lambda q(\lambda)P(a|x,\lambda)P(b|y,\lambda)$, where $P(a|x,\lambda)$ is the probability that Alice gets the outcome $a$ for the input $x$, determined by the variable $\lambda$, chosen from the distribution $q(\lambda)$ and similarly for $P(b|y,\lambda)$. Elements of $\textbf{p}$, which satisfy the LC relation form the set of local correlations $\mathcal{L}$. This set is closed, bounded, convex and forms a polytope. Certain correlations in quantum mechanics are not compatible with local correlations; this is known as Bell nonlocality. The elements of $\textbf{p}$ belong to the set of quantum correlations $\mathcal{Q}$ if,
$p(ab|xy)={\rm Tr}(\rho_{AB}M_{a|x}\otimes M_{b|y})$, where $M_{a|x}$ and $M_{b|y}$ are POVM elements of corresponding measurements. Set of quantum correlations is bounded and convex, but it is not a polytope as there are infinite number of extremal points. Also, this is not closed also as recently shown in the Ref. \cite{quantum_correlation0}.
Any behavior \textbf{p} is no-signaling $\mathcal{NS}$, if it satisfies the no-signaling constraints,
\begin{eqnarray}
\nonumber
&&\sum_b p(ab|xy)=\sum_b p(ab|xy'), \forall a,x,y,y'.\\
&&\sum_a p(ab|xy)=\sum_a p(ab|x'y), \forall a,x,y,y'.
\end{eqnarray}
No-signaling correlations also form a polytope, which consist of both local and nonlocal vertices. Both $\mathcal{L}$ and $\mathcal{Q}$ satisfy the no-signaling constraints, but there are $\mathcal{NS}$ correlations that do not satisfy locality and also do not belong to $\mathcal{Q}$. Any local behavior admits a quantum description and hence belongs to $\mathcal{Q}$. But there are quantum correlations that do not belong to $\mathcal{L}$. So, finally, we have $\mathcal{L}\subset\mathcal{Q}\subset \mathcal{NS}$, which is shown in the Fig. \ref{correlationfig}.
\begin{figure}[h]
\begin{center}
\includegraphics[scale=0.2]{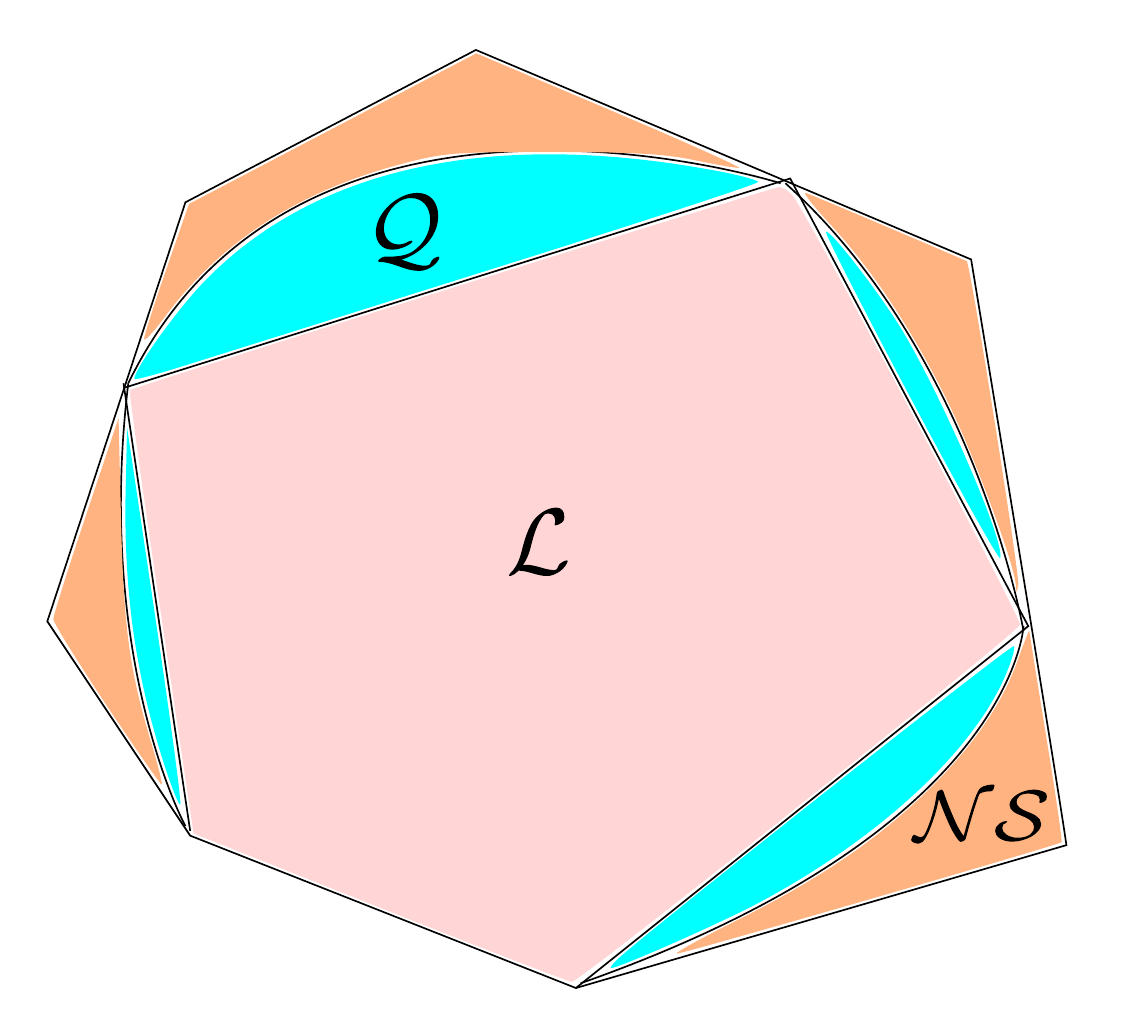}
\caption{ Schematic diagram of different type of correlations. }
\label{correlationfig}
\end{center}
\end{figure}
The set of quantum correlations is not as simple as it may appear from the schematic diagram. Recently, in \cite{quantum_correlation} authors have investigated the nontrivial geometry of a set of quantum correlations.
From hyperplane separation theorem \cite{hyperplane}, for each behavior \textbf{p} which is not the part of $\mathcal{L}$ or $\mathcal{Q}$ or $\mathcal{NS}$, there is a hyperplane that separates this \textbf{p} from the corresponding set. If the set is $\mathcal{L}$ then this is nothing but a Bell inequality. From the Fig. \ref{correlationfig}, it is evident that facet Bell inequality is the tightest or optimal Bell inequality for a set of local correlations. One can in principle construct Bell inequalities that are not facets of the local polytope, but these would not be optimal in the sense that there may be some quantum correlations which are nonlocal with respect to a facet Bell inequality, but do not violate the non-optimal one. So, it is always desirable to find facet Bell inequalities for a set of local correlations. In literature, facet Bell inequalities have been constructed for many scenarios \cite{brunnerRMP}, like for higher dimensions, different measurement settings, multipartite settings etc. As we have seen, one of the important features of a local polytope is that only local correlations are inside it. The set of quantum correlations contains the local polytope inside it but there are quantum correlations which are outside the local polytope  as seen from the Fig. \ref{correlationfig}. Therefore, some quantum correlations are expected to violate at least one of the facet inequalities of a given local polytope. From this point of view, it is of value to consider a local polytope with the smallest number of nontrivial facet inequalities. \\\\
As stated in the introduction, we first construct facet Bell inequalities for $\{[2 ~2] ~[2~ 2] ~[2]\}$ scenario.
For this case,  we have a local polytope of dimension $17$ with $32$ vertices (see \ref{appendixA}) in $V$ representation. By converting this $V$-representation to $H$-representation with the software cdd \cite{cdd} we obtained total of $48$ facet inequalities. Among $48$ inequalities, $32$ are just the positivity conditions for probabilities.  Remaining 16 inequalities are one of the four inequalities written below.  In this list, the left-hand side should be thought of as the expectation value of the observables.
\begin{eqnarray}
\nonumber
&&(A_2 B_2-A_2B_1-A_1B_2-A_1B_1)\\
\label{first}
&&+(A_2 B_2-A_2B_1-A_1B_2-A_1B_1)C_1-2C_1 \leq 2,\\[10pt]
\nonumber
&& (-A_2 B_2+A_2B_1+A_1B_2+A_1B_1)\\
\label{second}
&& +(-A_2 B_2+A_2B_1+A_1B_2+A_1B_1)C_1-2C_1 \leq 2,\\[10pt]
\nonumber
&& (A_2 B_2-A_2B_1-A_1B_2-A_1B_1)\\
\label{third}
&& +(-A_2 B_2+A_2B_1+A_1B_2+A_1B_1)C_1+2C_1 \leq 2,\\[10pt]
\nonumber
&& (-A_2 B_2+A_2B_1+A_1B_2+A_1B_1)\\
\label{fourth}
&& +(A_2 B_2-A_2B_1-A_1B_2-A_1B_1)C_1+2C_1 \leq 2.
\end{eqnarray}
In terms of the well known CHSH inequality, these four can be written more simply as,
\begin{eqnarray}
-I_{CHSH}-I_{CHSH}C_1-2C_1 & \leq 2, \\
 I_{CHSH}+I_{CHSH}C_1-2C_1 & \leq 2, \\ 
-I_{CHSH}+I_{CHSH}C_1+2C_1 & \leq 2, \\
 I_{CHSH}-I_{CHSH}C_1+2C_1 & \leq 2.
\end{eqnarray}
But, these four inequalities are equivalent. We can see that if we make the interchange of the indices as, $A_1\rightarrow A_2$,  $A_2\rightarrow -A_1$, $B_1\rightarrow B_2$, $B_2\rightarrow -B_1$ in the  first inequality (Eq. (\ref{first})), then it goes to the second inequality (Eq. (\ref{second})). Similarly, one can see that with this type of interchange all the above inequalities are equivalent. So, finally we have only one inequality. We will choose the form of second inequality (if not mentioned) to do the rest of the analysis.
 This inequality is equivalent to the lifted version of CHSH inequality (Eq. (2) of Ref. \cite{pironio_facet}) for more parties.
Now other than Charlie, one can choose either Alice or Bob doing one measurement and rest are doing two dichotomic measurements. For each case we get one facet Bell inequality. In this way, there are three inequalities, where in our previous paper we had six inequalities. These three inequalities are,
\begin{eqnarray}
\label{inchsh1}
&&I_1= I_{CHSH}+I_{CHSH}A_1-2A_1  \leq 2,\\
\label{inchsh2}
&&I_2 =I_{CHSH}+I_{CHSH}B_1-2B_1  \leq 2,\\
\label{inchsh3}
&&I_3= I_{CHSH}+I_{CHSH}C_1-2C_1 \leq 2.
\end{eqnarray}
In the following, we analyze  these facet Bell inequalities for different purposes. 


\section{Three-Qubit generalized GHZ states}
\label{generalized ghz}
In this section, we show that with the facet Bell inequalities, we can again have a violation for all generalized GHZ states like our previous paper's inequalities. We use average Von Neumann entropy over each bi-partition as a measure of entanglement. 
Below, we show that for the similar measurement settings used in our previous paper \cite{arpan}, all generalized GHZ states violate the facet Bell inequalities.
Let us consider the three-qubit generalized GHZ state,
\begin{equation}
\label{GGHZ}
\ket{GGHZ}=\alpha \ket{000}+\beta \ket{111}.
\end{equation}
Without loss of generality, for simplicity, we take $\alpha$ and $\beta$ to be real and positive numbers, as the method will be the same even if they are complex. 
Average Von Neumann entropy for generalized GHZ state as defined above over these bi-partitions is $-\alpha^2 \log_2 {\alpha^2}-\beta^2 \log_2 {\beta^2}$, which is also the entropy for each bi-partition for these states.
Now to see the violation by these states of the facet Bell inequality, let us take the following inequality, 
\begin{equation}\label{facet inequality}
I_B=I_{{CHSH}}+I_{{CHSH}}C_1-2C_1\leqslant 2,
\end{equation}
where $I_{{CHSH}}=A_1B_1+A_1B_2+A_2B_1-A_2B_2$. We choose $A_1=\sigma_z$, $A_2=\sigma_x$, $B_1=\cos\theta\sigma_x+\sin\theta\sigma_z$, $B_2=-\cos\theta\sigma_x+\sin\theta\sigma_z$ and $C_1=\sigma_x$. For the generalized GHZ state $\ket{GGHZ}=\alpha\ket{000}+\beta\ket{111}$, the expectation value of the operator $I_B$ is
\begin{equation}\label{facet inequality expectation}
\bra{GGHZ}I_B\ket{GGHZ}=2\sin\theta+4\alpha\beta\cos\theta.
\end{equation} 
As, $a\sin\theta+b\cos\theta\leqslant\sqrt{a^2+b^2}$, we have $\langle I_B \rangle_{\ket{GGHZ}}\leqslant 2\sqrt{1+4\alpha^2\beta^2}$ = $2\sqrt{1+ {{\cal C}^2}}$, where ${\cal C} = 2 {\alpha} {\beta}$ is nothing but the tangle \cite{tangle} of the generalized GHZ state.
The quantity
${\cal C}$ is also like concurrence for a two-qubit bipartite state. Maximum is achieved when we choose $\sin\theta=\frac{1}{\sqrt{1+4\alpha^2\beta^2}}$ and $\cos\theta=\frac{2\alpha\beta}{\sqrt{1+4\alpha^2\beta^2}}$.

Therefore, it is evident that as long as the state is entangled, {\it i.e.} $\alpha$ and $\beta$ are not zero, the generalized GHZ states will violate the facet Bell inequality. This proves our claim. 
We numerically optimize over all measurement settings and find that the maximum value of $I_B$ for the GHZ state is $2\sqrt{2}$. Hence, $2\sqrt{2}$ may be the maximum possible violation achieved by the GHZ state for the facet Bell inequality. For this measurement setting stated above,  we notice that the violation of a generalized GHZ state changes monotonically with entanglement. Hence, the more entangled a generalized GHZ state is, the more will be the violation of the facet Bell inequality for this particular measurement setting.\\
One question may now arise that for this particular measurement setting, we are getting the expression of violation which is a monotonic function of $\mathcal{C}$. If we choose other measurement setting, will this type of relation emerge? To answer this question,  let us consider a general measurement setting as below,
\begin{eqnarray}
&A_1=\sin\theta_{a1}\cos\phi_{a1}\sigma_x+\sin\theta_{a1}
\sin\phi_{a1}\sigma_y+\cos\theta_{a1}\sigma_z\nonumber,\\
&A_2=\sin\theta_{a2}\cos\phi_{a2}\sigma_x+\sin\theta_{a2}
\sin\phi_{a2}\sigma_y+\cos\theta_{a2}\sigma_z\nonumber,\\
&B_1=\sin\theta_{b1}\cos\phi_{b1}\sigma_x+\sin\theta_{b1}
\sin\phi_{b1}\sigma_y+\cos\theta_{b1}\sigma_z\nonumber,\\
&B_2=\sin\theta_{b2}\cos\phi_{b2}\sigma_x+\sin\theta_{b2}
\sin\phi_{b2}\sigma_y+\cos\theta_{b2}\sigma_z\nonumber,\\
&C_1=\cos\phi_{c1}\sigma_x+\sin\phi_{c1}\sigma_y\nonumber.
\end{eqnarray}
With these measurement settings we get
\begin{equation}\label{facet inequality expectation1}
\langle I_B \rangle_{\ket{GGHZ}}=X+\mathcal{C}Y,
\end{equation}
where, $X=\cos\theta_{a2}(\cos\theta_{b1}-\cos\theta_{b2})+\cos\theta_{a1}(\cos\theta_{b1}+\cos\theta_{b2})$, $Y=\cos(\phi_{a1}+\phi_{b1}+\phi_{c1})\sin\theta_{a1}
\sin\theta_{b1}+\cos(\phi_{a2}+\phi_{b1}+\phi_{c1})\sin\theta_{a2}
\sin\theta_{b1}+\cos(\phi_{a1}+\phi_{b2}+\phi_{c1})\sin\theta_{a1}
\sin\theta_{b2}-\cos(\phi_{a2}+\phi_{b2}+\phi_{c1})\sin\theta_{a2}
\sin\theta_{b2}$ and $\mathcal{C}=2\alpha\beta$.
From the above relation, it is clear that for fixed values of $X$ and $Y$, the amount of violation is again monotonic in $\mathcal{C}$. It is worth mentioning that this monotonic behavior is evident if one restricts to rank-1 projective measurements.
So, no matter what the measurement settings, we will get more violation for a more entangled state, as long as we use same measurement settings for the states.
\subsection{Comparison with Mermin inequality}
Mermin inequality \cite{mermin} can also track the entanglement, {\it i.e.} violation of Mermin inequality will be more for more entangled generalized GHZ states.
Mermin inequality is
\begin{equation}\label{mermin inequality}
I_{\mbox{M}}=A_1B_1C_2+A_1B_2C_1+A_2B_1C_1-A_2B_2C_2\leq 2.
\end{equation}
In this case if we choose the same general measurement settings as described above with $C_2=\cos\phi_{c2}\sigma_x+\sin\phi_{c2}\sigma_y$. The expectation value of the operator $I_{\mbox{M}}$ for the generalized GHZ state is
\begin{eqnarray}\label{mermin inequality expectation}
\nonumber
\langle I_{\mbox{M}}\rangle_{\ket{GGHZ}} &=&\mathcal{C}\Big(
\cos(\phi_{a1}+\phi_{b1}+\phi_{c2})sin\theta_{a1}
\sin\theta_{b1}+\\
&&\cos(\phi_{a2}+\phi_{b1}+\phi_{c1})\sin\theta_{a2}\sin\theta_{b1}+\nonumber\\
&&+\cos(\phi_{a1}+\phi_{b2}+\phi_{c1})sin\theta_{a1}
\sin\theta_{b2}-\nonumber\\
&&\cos(\phi_{a2}+\phi_{b2}+\phi_{c2})sin\theta_{a2}
\sin\theta_{b2}\Big).
\end{eqnarray}
So, the expectation value of the Mermin operator is again a monotonic function of $\mathcal{C}$. But the problem is that it does not show violation for the whole range of generalized GHZ states. So, for those states which do not violate Mermin inequality, this relation between entanglement and nonlocality has no meaning. But this relation can be used to measure the entanglement.

\subsection{More Violation by a non-maximally entangled state}
Unlike our previous inequalities, which are violated maximally by the GHZ state by an amount $2\sqrt{2}$, the facet Bell inequalities are violated
more by other genuinely entangled states. Quantum violation of lifted Bell inequalities is recently explored in \cite{jeba}. It was found that only the tensor product of Bell states and some auxiliary states could give the maximal violation. 
One very simple example is the W state. Numerically, we have found that the W state gives the maximum violation of 3.105 for the inequality, where Charile makes one measurement. Obviously, there is no ordering of violation of the facet Bell inequality according to the entanglement within the W class. Like the state $\sqrt{1/6}\ket{001}+\sqrt{3/6}\ket{010}+\sqrt{2/6}\ket{100}$ has average entropy 0.856 and violation of 3.33.  And $\sqrt{1/10}\ket{001}+\sqrt{4/10}\ket{010}+\sqrt{5/10}\ket{100}$ has average entropy 0.813 and violation 3.475.  Ordering is valid only for generalized GHZ states, not for the whole GHZ class.
 Not only that, there are states within GHZ class, which violate the facet Bell inequality more than the conventional GHZ state. Like the state $\ket{\psi}=\sqrt{22/50}\ket{000}+\sqrt{3/50}\ket{100}+\sqrt{2/50}\ket{101}+\sqrt{21/50}\ket{110}+\sqrt{2/50}\ket{111}$
has maximum expectation value  $3.377$ (found numerically) and  also belongs to the GHZ class.  From Ref. \cite{lianglifted} one can compare the maximum quantum bound for the facet Bell inequalitiy, which is $\approx 3.657$. 
 For three-qubit systems, the GHZ-state can be considered to be the maximally entangled state. In this case, the subsystems are maximally mixed.
Furthermore, for several communication protocols, the GHZ state is a task-oriented maximally entangled state \cite{task}.
But we see, that a facet Bell inequality is not maximally violated by this state. Non facet inequalities like in reference \cite{arpan}, and Mermin inequalities are violated maximally by the GHZ-state.
\subsection{Three-qubit pure bi-separable states and genuinely entangled states}
The three facet Bell inequalities explore the entanglement of three types of bi-separable pure states like our previous inequalities \cite{arpan}. For example, the state which is separable in $1-23$ bi-partition will violate that facet Bell inequality, which can explore the entanglement 
between the second and the third qubit. So in this case, the inequality Eq. (\ref{inchsh1}), i.e $I_{CHSH}+I_{CHSH}A_1-2A_1  \leq 2$ will be violated. Similarly, the other two types of bi-separable states will violate the other two inequalities.
But we can not distinguish between bi-separable and genuinely entangled pure states like our previous set of inequalities. Because we had six inequalities in the previous paper and a bi-separable state would violate exactly two inequalities with the same amount of optimal violation. But in the case of facet Bell inequalities, a bi-separable states violates only one out of the three, and that optimal violation may be exhibited by some genuinely entangled state also. So, by a violation, we can not say whether a state is a bi-separable pure state or a genuinely entangled pure state.\\
In the Ref. \cite{popescuall}, authors demonstrated the nonlocality of a $n$-qubit entangled state, by showing that there exists a local projective measurement by $n-2$ parties, such that the resulting state of the remaining two parties violate CHSH inequality. This fact was further elaborated in the Ref. \cite{cavalall}, where the authors constructed generalized Bell inequalities for $n$-partite systems using that it is always possible to construct a Bell inequality that is violated by a $n$-partite state from a Bell inequality, which is violated by a post measurement state of $n-k$ parties resulted from the local projective measurements by $k$ parties. This is exactly the idea of lifted Bell inequalities, which is equivalent to our facet inequalities. As a result, these inequalities will be violated by any three-qubit pure entangled state. 
So, a genuinely entangled pure three-qubit state will also violate the inequalities, but these inequalities are not suitable for distinguishing between genuinely entangled pure three-qubit states and bi-separable pure three-qubit states. 

\section{Mixed state scenario}
Mixed states present different challenges. There is a phenomenon of hidden nonlocality \cite{popescuhidden,gisinfil,hirsch,bowles,palazuelos,liang}. We have the modest goal to find
where the facet Bell inequalities of this paper may be more useful. We consider a few noisy states, like noisy GHZ states and noisy W states with both white and colored noise, to see whether any advantages exist for the facet Bell inequalities over the Mermin inequality for mixed states. First, we will take a Werner-like state for three qubits, which is a GHZ state with white noise.
\begin{equation}
NoisyGHZ=p\ket{GHZ}\bra{GHZ}+\frac{(1-p)}{8}\mathds{1},
\end{equation}
where,
\begin{eqnarray}
\nonumber
&&\mathds{1}=\ket{\psi_0^+}\bra{\psi_0^+}
+\ket{\psi_0^-}\bra{\psi_0^-}+\ket{\psi_1^+}
\bra{\psi_1^+}
+\ket{\psi_1^-}\bra{\psi_1^-}\\
&&+\ket{\psi_2^+}\bra{\psi_2^+}
+\ket{\psi_2^-}\bra{\psi_2^-}+\ket{\psi_3^+}\bra{\psi_3^+}+
\ket{\psi_3^-}\bra{\psi_3^-},
\end{eqnarray}
and,
\begin{eqnarray}
&&\ket{\psi_0^+}=\ket{GHZ}=\sqrt{1/2}(\ket{000}+\ket{111}), \\
&& \ket{\psi_0^-}=\sqrt{1/2}(\ket{000}-\ket{111}), \\
&& \ket{\psi_1^+}=\sqrt{1/2}(\ket{010}+\ket{101}), \\
&& \ket{\psi_1^-}=\sqrt{1/2}(\ket{010}-\ket{101}), \\
&& \ket{\psi_2^+}=\sqrt{1/2}(\ket{100}+\ket{011}), \\
&& \ket{\psi_2^-}=\sqrt{1/2}(\ket{100}-\ket{011}), \\
&& \ket{\psi_3^+}=\sqrt{1/2}(\ket{110}+\ket{001}), \\
&& \ket{\psi_3^-}=\sqrt{1/2}(\ket{110}-\ket{001}).
\end{eqnarray}

For this noisy GHZ state, we have numerically obtained the optimal expectation value of the  facet Bell operator  and Mermin operator for the whole range of $p$ ($0\leq p \leq 1$) and plotted them in Fig. \ref{werplot}.  We will call the value of $p$ as the critical value $p_c$, such that the inequalities get violated for $p>p_c$. 
 For noisy GHZ states $p_c=0.707$ for facet Bell inequality and $p_c=0.500$ for the Mermin inequality.
\begin{figure}[H]
\begin{center}
\includegraphics[height=4.0cm,width=6.4cm]{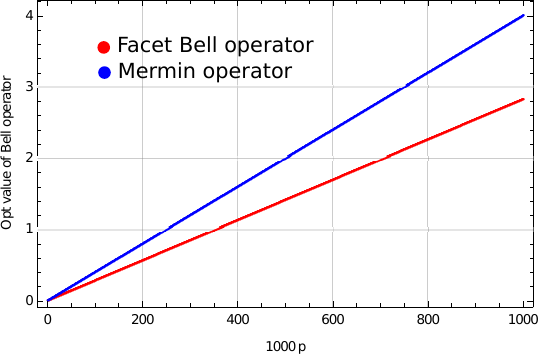}
\caption{ Maximum expectation value of the our Bell operator and  Mermin operator for noisy GHZ states vs $p$ plot.}
\label{werplot}
\end{center}
\end{figure}
So, for the noisy GHZ states, our facet Bell inequality presents no advantage. One of the reasons for this is that, Mermin inequality is optimally constructed for GHZ states, giving a violation of $4$, whereas, the facet Bell inequality gives only $2\sqrt{2}$ for GHZ states.
\begin{figure}[H]
\begin{center}
\includegraphics[height=4.0cm,width=6.4cm]{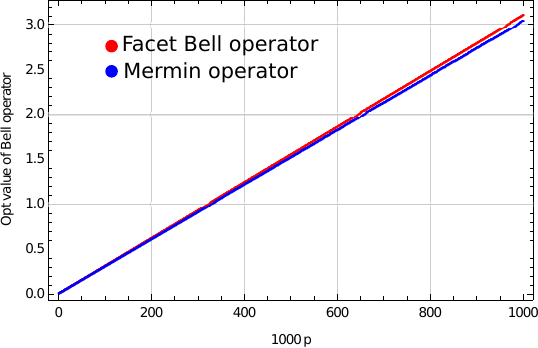}
\caption{ Maximum expectation value of our Bell operator and the Mermin operator for noisy W states vs $p$ plot.}
\label{werplotmer}
\end{center}
\end{figure}
\noindent Let us now consider noisy W states to analyze the same thing (see Fig. \ref{werplotmer}) taking,
\begin{equation}
NoisyW=p\ket{W}\bra{W}+\frac{(1-p)}{8}\mathds{1}.
\end{equation}

In this case, $p_c=0.644$ for facet Bell inequality and $p_c=0.656$ for Mermin inequality. So, in this case, the facet Bell inequality gives a slight advantage over the Mermin inequality. If we take colored noise and different noisy states, we can see that sometimes our inequality has an advantage over the Mermin.
Similarly, we can take colored noise and do the same analysis as before. In the following, we provide the Table \ref{noisytable} listing the results we have obtained numerically.

\begin{table}[ht]
\centering
\begin{tabular}{|c |c|c|c|}
\hline

&\multicolumn{2}{|r|}{Critical value $p_c$} \\\cline{2-3}
{States}  &    Facet &   Mermin \\
\hline
$p\ket{W}\bra{W}+\frac{(1-p)}{8}\mathds{1}$ &0.644&0.656\\
\hline
$p\ket{W1}\bra{W1}+\frac{(1-p)}{8}\mathds{1}$ &0.600&0.670\\
\hline
$p\ket{GHZ}\bra{GHZ}+\frac{(1-p)}{8}\mathds{1}$ &0.707&0.500\\
\hline
$p\ket{GGHZ_0}\bra{GGHZ_0}+\frac{(1-p)}{8}\mathds{1}$ & 0.855 & 0.913\\
\hline
$p\ket{GHZ}\bra{GHZ}+\frac{(1-p)}{5}col$ &0.633&0.375\\
\hline
\end{tabular}
\caption{Critical value $p_c$ for different noisy states.}
\label{noisytable}
\end{table}

\begin{eqnarray}
&& \ket{GGHZ_0}=\sqrt{45/49}\ket{000}+\sqrt{4/49}\ket{111}, \\
&&col=\ket{\psi_0^+}\bra{\psi_0^+}+\ket{\psi_1^+}\bra{\psi_1^+}+\ket{\psi_1^-}\bra{\psi_1^-}+\ket{\psi_2^+}\bra{\psi_2^+}+\ket{\psi_2^-}\bra{\psi_2^-},\\
&&\ket{W}=\sqrt{1/3}\ket{001}+\sqrt{1/3}\ket{010}+\sqrt{1/3}\ket{100},\\
&&\ket{W1}=\sqrt{1/6}\ket{001}+\sqrt{2/6}\ket{010}+\sqrt{3/6}\ket{100}.
\end{eqnarray}
In the above, we have taken $col$ to be colored noise, and $\ket{W1}$ is a W class state. 
Evidently, our inequality gives advantages for noisy W and $\ket{W1}$ states. For noisy GHZ states, the Mermin inequality is better. If we replace GHZ state by a generalized GHZ state (in the form $\cos\theta\ket{000}+\sin\theta\ket{111}$), for the cases starting from the close vicinity of the parameter range $\theta= 15^{\circ}\hspace{1mm} i.e.,\hspace{1mm} \sin\theta \approx 0.25$  (below which Mermin does not get violated) our inequality is more effective than Mermin's. From the table it is evident that when $\sin\theta=\sqrt{4/49}\approx 0.286$, the noisy state $\ket{GGHZ_0}$ violates Mermin when $p>p_c=0.913$, whereas for facet Bell inequality $p_c=0.855$. Obviously, in those regions, our inequality is advantageous, because they are violated by all generalized GHZ states, {\it i.e} GGHZ states. One can, in principle, check for other mixed states. We have analyzed the noisy ones because they are experimentally relevant. Whenever one tries to prepare a GHZ or a W state in a lab, unavoidable noises add up, making the states noisy.

\section{Extension to multipartite scenario}
In this section, we will extend the previous facet Bell inequalities to more than three parties. First, we will deal with a four-qubit scenario and then with five qubits. After that results will be generalized for $n$ qubits, where $n\geq 3$. In all these scenarios, we will be restricting our calculations for the situations, where two parties are making two dichotomic measurements and the rest are making only one dichotomic measurement. For this particular scenario, we will find nontrivial facets of the local polytope. This scenario is minimal scenario. Let's start with four qubits.
\subsection{Four qubits scenario}
For this case, specifically, for the $\{[2 ~2] ~[2~ 2] ~[2]~[2]\}$ scenario,  we have a $35$ dimensional local polytope with $64$ vertices. We again convert this $V$-representation of the polytope to the $H$-representation using the software cdd \cite{cdd} and obtain a total of $96$ facets. Out of which $64$ facets are just the positivity conditions on the probabilities. So, we get $32$ nontrivial facet inequalities. But, interestingly, these $32$ inequalities are just the variants of one single inequality, up to the relabeling of indices. So, like the three-qubit scenario, we again get only one single inequality,
\begin{eqnarray}
\nonumber
&(-2+A_1(B_1+B_2)+A_2(B_1-B_2))(1+C_1)(1+D_1)\leq 0.
\end{eqnarray}
All the $32$ facet inequalities are equivalent to this inequality up to the relabeling of indices. The form of this inequality is very similar to the inequality for the three-qubit case. Because one can write the inequality given by the Eq. (\ref{inchsh3}) as,
\begin{equation}
\nonumber
(-2+A_1(B_1+B_2)+A_2(B_1-B_2))(1+C_1)\leq 0,
\end{equation}
which has similar structure like that for four-qubit inequality, except the extra party is denoted by $D$. Next, we will explore whether the five-qubit case also has the similar structure.
\subsection{Five qubits or more}
In the scenario $\{[2 ~2] ~[2~ 2] ~[2]~[2]~[2]\}$, we have a  $71$ dimensional local polytope with $128$ vertices. Converting from the $V$ representation to $H$ representation for this local polytope, we obtain a total of $192$ facets. Out of which, $128$ inequalities are just the positivity conditions for the probabilities. The remaining $64$ inequalities again give only one non-trivial inequality up to the relabelling of the indices. 
\begin{eqnarray}
\nonumber
&(-2+A_1(B_1+B_2)+
A_2(B_1-B_2))(1+C_1)(1+D_1)(1+E_1)\leq 0.
\end{eqnarray}
Again for the five-qubit case, we have the same structure of the inequality as that for  three- and four-qubit cases, with the addition of a new term for the party $E$. So, after exploring these three cases extensively, we can generalize this structure to more qubits. For $n$ number of qubits, we can generalize the structure as,
 \begin{eqnarray}
\label{nquibit}
\nonumber
&(-2+A_1(A_2+A'_2)+A'_1(A_2-A'_2))
(1+A_3)(1+A_4)...(1+A_n)\leq 0,
\end{eqnarray}
where $A_1$ and $A'_1$ are the two measurement choices for the party $1$ and so on. If we just expand this we will get,
\begin{eqnarray}
\label{nquibit}
\label{nquiex}
\nonumber
&(A_1(A_2+A'_2)+A'_1(A_2-A'_2))(1+A_3)...(\mathds{1}+A_n)\\
&-(2A_3+2A_4+...+2A_3A_4+...+2A_3A_4..A_n)\leq 2.
\end{eqnarray}
So, the facet Bell inequalities have very simple and intuitive structure. 
We can permute the parties that make two dichomotic measurements to obtain the complete set.  We now show that all $n$-qubit generalized GHZ state violate this $n$-qubit facet Bell inequality.
\subsection{Violation by $n$-qubit GGHZ state}
{As explained above, using the results from Ref. \cite{popescuall},  this multipartite facet Bell inequality will also be violated by any $n$-qubit entangled pure state. 
Here, we will show that for the similar measurement settings the $n$-qubit facet Bell inequality for $n$ qubits will be violated by the generalized GHZ states for the whole parameter range. To show this, we take the $n$-qubit generalized GHZ state to be, $\ket{GGHZn}=\alpha\ket{00..0_n}+\beta\ket{11..1_n}$ and the similar measurement settings as the three-qubit scenario, i.e we choose, $A_1=\sigma_z$, $A'_1=\sigma_x$, $A_2=\cos\theta\sigma_x+\sin\theta\sigma_z$, $A'_2=-\cos\theta\sigma_x+\sin\theta\sigma_z$ and all other measurement settings to be $\sigma_x$, i.e $A_3=\sigma_x$, $A_4=\sigma_x$,...,$A_n=\sigma_x$. Now for these measurement settings the expectation value of the facet Bell operator given by the Eq. (\ref{nquiex}) is $(2\sin\theta+4\alpha\beta\cos\theta)$, which is exactly equal to the previously obtained expectation value for the three-qubit scenario. So, the generalized GHZ state will violate the $n$-qubit facet Bell inequality for all the range of parameters, giving the violation of $2\sqrt{1+4\alpha^2\beta^2}$ for this measurement settings.

\section{Conclusion}

In this paper, we have considered a specific measurement scenario. This scenario may be thought of as the
{\it minimal} scenario that involves multipartite correlations.  In this scenario, there are two dichotomic 
measurement settings for two parties and one dichotomic measurement setting for each of the remaining
 parties. Interestingly, there is just one facet Bell inequality (up to permutation of parties) 
 for $n$ qubits. 
This is like the two-qubit scenario where only CHSH inequality is the facet Bell inequality. 
This suggests that we need only one facet Bell inequality that uses multipartite correlations 
to detect the nonlocality of a multipartite
state. This gives significant advantage over the other scenarios.
In this scenario, we first constructed facet Bell inequalities for a three-qubit system. This was motivated by our previous work \cite{arpan}. Then, we showed that the three facet Bell inequalities give similar advantages like our previous inequalities \cite{arpan}. However, the  facet Bell inequalities are now not violated maximally by the GHZ states, which can be considered as maximally entangled three-qubit state. We then computed the facets for four and five qubits in the {\it minimal} scenario. We found that each of these two cases again give only one non-trivial facet inequality up to the relabelling of indices as expected from the results of the Ref. \cite{pironio_facet}. We then extended our results to $n$ parties and showed that the $n$-qubit facet Bell inequality is violated by all $n$-qubit generalized GHZ states.
We have compared the {\it minimal} scenario facet three-qubit inequalities with Mermin inequalities and also analyzed some cases of mixed states, including noisy GHZ and W states.
We have demarcated where these facet Bell inequalities present advantages. Inequalities in this paper can be tested experimentally as our previous ones \cite{experiment}. Notably, in Ref. \cite{lvw},
similar inequalities were discussed in a different context, and in Ref. \cite{acin_new} authors used the lifted Bell inequalities to detect genuine multipartite nonlocal correlations.

\section*{Acknowledgement}
P.A. acknowledges the support from the Department of Science and Technology, India, through the project DST/ICPS/QuST/Theme-1/2019.\\\\



\appendix
\section{Appendix}
\label{appendixA}
In our Bell test scenario, we have three parties with two dichotomic measurements for two parties and one dichotomic measurement for the other party. In a Bell test we usually measure the joint outcome probabilities i.e., $P(abc|xyz)$. Here $x,y\in\{0,1\}$ are the measurement settings for Alice and Bob respectively and $a,b\in\{0,1\}$ are the corresponding outcome for Alice and Bob respectively. As here Charlie is doing one measurement so $z=0$ and $c\in\{0,1\}$. Therefore, there are total of 32 joint probabilities. But not all of them are independent. No-signaling and normalization conditions constrain the number of independent joint probabilities and determine the dimension of the probability space \cite{symmetric}, which is $[(m_1(d-1)+1).(m_2(d-1)+1).(m_3(d-1)+1)]-1$. In our case $m_1=2,m_2=2,m_3=1,d=2$, which gives the dimension to be $17$. Now the conditions of local correlations will determine the vertices. First, we have to choose the parametrization for this $17$ dimensional space.
We choose the following parametrization.
\begin{eqnarray}
P&=&[P(a_0),P(a_1),P(b_0),P(b_1),P(c_0),P(a_0b_0),P(a_0b_1),\nonumber\\
&&P(a_1b_0),P(a_1b_1), P(a_0c_0),P(a_1c_0),P(b_0c_0),P(b_1c_0),\nonumber\\&&P(a_0b_0c_0),P(a_0b_1c_0),
P(a_1b_0c_0), P(a_1b_1c_0)],\label{joint probability space}
\end{eqnarray}
where $P(a_x)=P(0|x)$, $P(b_y)=P(0|y)$, $P(c_z)=P(0|z)$, $P(a_xb_y)=P(00|xy)$ and $P(a_xb_yc_z)=P(000|xyz)$. So this 17 dimensional polytope consists of 32 extremal points or vertices. This polytope has been described using the V-representation. One can find the facets of this polytope using some standard algorithm. The number of facets for this polytope is 48.
Now, we can write the probabilities in terms of expectation values, like $P(a_x)=1/2(1+\langle a_x \rangle)$ and similarly for the joint probabilities. By this substitution of expectation values in place of probability distributions, we can write the facets as following,
\begin{eqnarray}
&&(1+A_x)(1+B_y)(1+C_z)\geq 0,\label{correlation1} \\
&&(1+A_x)(1+B_y)(1-C_z)\geq 0,\label{correlation2} \\
&&(1+A_x)(1-B_y)(1+C_z)\geq 0,\label{correlation3}\\
&&(1-A_x)(1+B_y)(1+C_z)\geq 0,\label{correlation4} \\
&&(1+A_x)(1-B_y)(1-C_z)\geq 0,\label{correlation5}\\
&&(1-A_x)(1+B_y)(1-C_z)\geq 0,\label{correlation6}\\
&&(1-A_x)(1-B_y)(1+C_z)\geq 0,\label{correlation7} \\
&&A_x(1-B_y)(1-C_z)+B_y(1-C_z)+C_z\leq 1,\label{correlation8} \\
&&[-2+A_0(B_0-B_1)-A_1(B_0+B_1)](1+C_0)\leq 0,\label{correlation9} \\
&&[2+A_1(-B_0+B_1)+A_0(B_0+B_1)](1+C_0)\geq 0,\label{correlation10}\\
&&[2+A_0(B_0-B_1)+A_1(B_0+B_1)](1+C_0)\geq 0,\label{correlation11} \\
&&[2+A_1(B_0-B_1)+A_0(B_0+B_1)](1+C_0)\geq 0,\label{correlation12} \\
&&[-2+A_0(B_0-B_1)+A_1(B_0+B_1)](1+C_0)\leq 0,\label{correlation13}
\end{eqnarray}
\begin{eqnarray}
&&[-2+A_1(B_0-B_1)+A_0(B_1+B_1)](1+C_0)\leq 0,\label{correlation14} \\
&&[2+A_0(B_0-B_1)-A_1(B_0+B_1)](1+C_0)\geq 0,\label{correlation15} \\
&&[-2+A_1(-B_0+B_1)+A_0(B_0+B_1)](1+C_0)\leq 0,\label{correlation16}\\
&&[A_0(B_0-B_1)-A_1(B_0+B_1)](1-C_0)+2C_0\leq 2,\label{correlation17} \\
&&[A_1(-B_0+B_1)+A_0(B_0+B_1)](-1+C_0)+\nonumber\\&&2C_0\leq 2,\label{correlation18} \\
&&[A_0(B_0-B_1)+A_1(B_0+B_1)](-1+C_0)+\nonumber\\&&2C_0\leq 2,\label{correlation19} \\
&&[A_1(B_0-B_1)+A_0(B_0+B_1)](-1+C_0)+\nonumber\\&&2C_0\leq 2,\label{correlation20}\\
&&[-2+A_0(B_0-B_1)+A_1(B_0+B_1)](-1+C_0)\geq 0,\label{correlation20} \\
&&[-2+A_1(B_0-B_1)+A_0(B_0+B_1)](-1+C_0)\geq 0,\label{correlation20} \\
&&[2+A_0(B_0-B_1)-A_1(B_0+B_1)](-1+C_0)\leq 0,\label{correlation20} \\
&&[-2+A_1(-B_0+B_1)+A_0(B_0+B_1)](-1+C_0)\geq 0,\label{correlation20}
\end{eqnarray}
{\color{red} First 32 inequalities (Eq. (\ref{correlation1}) - Eq.(\ref{correlation8})) are just the positivity conditions. Remaining 16 inequalities one of the four inequalities (Eq. (\ref{first}) - Eq. (\ref{fourth})) written in the main text.}
Similarly, we have computed the facets for four and five {\color{red}parties} and found that they have similar structure.\\

\begin{thebibliography}{}

\bibitem{bell1964} J. S. Bell,  {\it Physics} {\bf 1}, 195 (1964).


\bibitem{EPR}
A. Einstein,  B. Podolsky, and N. Rosen,
{\it Phys. Rev.} {\bf 47}, 777 (1935).

\bibitem{werner}
 R. F. Werner, {\it Phys. Rev. A} {\bf 40}, 4277 (1989).

\bibitem{brunnerRMP}
N. Brunner {\it et al}.,  {\it Rev. Mod. Phys.},  {\bf 86}, 419  (2015).


\bibitem{chsh}
J. F. Clauser {\it et al}., {\it Phys. Rev. Lett.} {\bf 23}, 880 (1969).


\bibitem{tsirelson}
B. S. Tsirelson, {\it Lett. Math. Phys.}  {\bf 4}, 93 (1980);
B. S. Tsirelson, {\it J. Sov. Math.}  {\bf 36} , 557 (1987).

\bibitem{sli}
C. {\'S}liwa, {\it Phys.  Lett. A} {\bf 317}, 165 (2003).

\bibitem{mermin}
N. D. Mermin, {\it Phys. Rev. Lett.}  {\bf 65}, 1838 (1990)

\bibitem{arpan}
A. Das, C. Datta, P. Agrawal, {\it Phys. Lett. A} {\bf 381},  3928 (2017)

\bibitem{ardehali}
M. Ardehali,  {\it Phys. Rev. A} {\bf 46}, 5375 (1992)

\bibitem{bk}
A. V.  Belinskii  and  D. N.  Klyshko,  {\it Phys.  Usp.}  {\bf 36}, 653 (1993).

\bibitem{gisinp}
N. Gisin, H. Bechmann-Pasquinucci, {\it Phys.  Lett. A} {\bf 246}, 1 (1998).

\bibitem{gen}
V. Scarani and N. Gisin, {\it J. Phys. A}
{\bf 34},
6043 (2001).

\bibitem{all}
R. F. Werner and M. M. Wolf,
{\it Phys. Rev. A}
{\bf 64}, 032112 (2001);
M. \.{Z}ukowski and \v C. Brukner, {\it Phys. Rev. Lett.} {\bf 88}, 210401 (2002).

\bibitem{nonvio}
M. \.{Z}ukowski, \v {C}. Brukner, W. Laskowski, and M.
Wie\' sniak, {\it Phys. Rev. Lett.} {\bf 88}, 210402 (2002).

\bibitem{degenerate}
P-S. Lin {\it et al}.,{\it Phys. Rev. A} {\bf 99}, 062338 (2019).

\bibitem{moremeasure}
W. Laskowski {\it et al}., {\it Phys. Rev. Lett.} {\bf 93}, 200401 (2004).

\bibitem{uffnik}
M. Seevinck, J. Uffink, {\it Phys. Rev. A} {\bf 65}, 012107 (2002).

\bibitem{nsvet}
D. Collins {\it et al.}, {\it Phys. Rev. Lett.} {\bf 88}, 170405 (2002).

\bibitem{nagata1}
K. Nagata, M. Koashi, and N. Imoto, {\it Phys. Rev. Lett.} {\bf 89}, 260401 (2002).

\bibitem{yu1}
S. Yu {\it et al.}, {\it Phys. Rev. Lett.} {\bf 90}, 080401 (2003).

\bibitem{lianghidden}
T. J. Barnea {\it et al.}, {\it Phys. Rev. A} {\bf 88}, 022123 (2013).

\bibitem{collins2}
D. Collins, and N. Gisin, {\it J. Phys. A: Math. Gen.} {\bf 37}, 1775 (2004).

\bibitem{pironio_facet}
S. Pironio, {J. Math. Phys.} {46}, 062112 (2005).


\bibitem{polytope}
G. M. Ziegler, {\it Lectures notes on polytope}, {\bf Springer} (1994).

\bibitem{quantum_correlation0}
W. Slofstra, {\it arXiv} {\bf 1703.08618} (2017).

\bibitem{quantum_correlation}
K. T. Goh {\it et al.}, {\it Phys. Rev. A}  {\bf 97}, 022104 (2018).

\bibitem{hyperplane}
S. P. Boyd and L. Vandenberghe, {\it Convex Optimization}, {\bf Cambridge University Press} (2004).

\bibitem{cdd}
https://www.inf.ethz.ch/personal/fukudak/cdd{\_}home/

\bibitem{tangle} V. Coffman, J. Kundu, and W. K. Wootters, {\it Phys. Rev. A} {\bf 61} 052306 (2000)


\bibitem{jeba}
C. Jebarathinam {\it et al}., {\it Phys. Rev. Research} {\bf 1}, 033073 (2019).

\bibitem{popescuall}
S. Popescu and D. Rohrlich, {\it Phys. Lett. A} {\bf 166}, 293 (1992).

\bibitem{cavalall}
D. Cavalcanti {\it et al}., {\it Nat. Commun.} {\bf 2}, 2184 (2011).


\bibitem{lianglifted}
J. Vallins, A. B. Sainz, and Y. C. Liang, {\it Phys. Rev. A} {\bf 95}, 022111 (2017).

\bibitem{task}
P. Agrawal and B. Pradhan, {\it J. Phys. A: Math. Theor.} {43}, 235302 (2010).


 



\bibitem{popescuhidden}
S. Popescu, {\it Phys. Rev. Lett.} {\bf 74}, 2622 (1995).

\bibitem{gisinfil}
N. Gisin, {\it Phys. Lett. A} {\bf 210}, 151 (1996).

\bibitem{hirsch}
F. Hirsch {\it et al}., {\it Phys. Rev. Lett.} {\bf 111}, 160402 (2013).

\bibitem{bowles}
J. Bowles {\it et al.}, {\it Phys. Rev. Lett.} {\bf 116}, 130401 (2016).

\bibitem{palazuelos}
C. Palazuelos, {\it Phys. Rev. Lett.} {\bf 109}, 190401 (2003).

\bibitem{liang}
Y.-C. Liang, L. Masanes, and D. Rosset, {\it Phys. Rev. A} {\bf 86}, 052115 (2012).


\bibitem{experiment}
J- Q. Zhao {\it et al}., {\it Phys. Lett. A} {\bf 382},  1214 (2017)

\bibitem{lvw}
W. Laskowski, T. Vertesi, and M. Wiesniak,  {\it J. Phys. A: Math. Theor.} {\bf 48}, 465301 (2015).

\bibitem{acin_new}
F. J. Curchod, M. L. Almeida,  and A. Acin, {\it New J. Phys.} {\bf 21}, 023016 (2019).

\bibitem{symmetric}
J. D. Bancal, N. Gisin and S. Pironio, {\it J. Phys. A: Math.
Theor.} {\bf 43}, 385303 (2010).

\end{thebibliography}
\end{document}